\documentstyle[preprint,aps]{revtex}

\begin{document}
\draft
\baselineskip 24 true pt
\vsize=11.0 true in \voffset=.2 true in
\hsize=6.5 true in \hoffset=.2 true in

\newpage
\begin{center}
{\Large{\bf First excited state calculation using 
different phonon bases for the two-site Holstein model }}\\
\end{center}
\vskip 1.0cm
\begin{center}
 Jayita Chatterjee\footnote{Electronic Address: moon@cmp.saha.ernet.in} 
 and A. N. Das\footnote{Electronic Address: atin@cmp.saha.ernet.in}
\end{center}
\vskip 0.50cm
\begin{center}

 {\em Saha Institute of Nuclear Physics \\
1/AF Bidhannagar, Calcutta 700064, India}\\

\end{center}

\vskip 1.0cm

PACS No.71.38. +i, 63.20.kr  
\vskip 1.0cm
\begin{center}
{\bf Abstract}
\end{center}
\vskip 0.5cm
 
  The single-electron energy and static charge-lattice 
 deformation correlations have been calculated for the first 
 excited state of a two-site Holstein model within perturbative 
expansions using different standard phonon bases obtained through 
Lang-Firsov (LF) transformation, LF with squeezed phonon states, 
modified LF, modified LF transformation 
with squeezed phonon states, and also within weak-coupling 
perturbation approach. Comparisons of the convergence of the perturbative 
expansions for different phonon bases reveal that modified LF approach 
works much better than other approaches for major range 
of the coupling strength.
\newpage
\begin{center}
{\bf I. INTRODUCTION}   
\end{center}
\vskip 0.3cm 

The simplest model to study the nature and properties of polarons 
as a function of electron-phonon ($e$-ph) interaction is the Holstein 
model \cite {Hol} which consists of tight binding electrons coupled 
through a site diagonal interaction term to 
dispersionless phonons. The ground state properties of this model 
has been extensively studied during recent years. In the 
antiadiabatic limit the confinement 
of the lattice deformation around the charge carrier for large $e$-ph 
interaction gives rise to 
small polarons whose nature and dynamics is generally studied using 
the Lang-Firsov (LF) method based on the canonical LF transformation 
\cite {LF}. For weak and intermediate $e$-ph coupling the LF method 
is not appropriate. For that region the importance and superiority 
of the modified LF (MLF) and the MLF with squeezed transformation (MLFS) 
over the LF method have been pointed out in previous 
works \cite {LS,DS,DC}. 
The results of exact diagonalization studies 
\cite {RT} of a two-site Holstein model indicate the failure 
of the standard (zero phonon averaging) classical LF approach 
even in the strong coupling antiadiabatic limit and the authors of 
Ref. \cite {RT} asserted that perturbation approach 
within the LF scheme is not meaningful. 
Marsiglio \cite {Mar} studied the Holstein model in one 
dimension with one-electron up to 
16 site lattices using numerical diagonalization technique 
and concluded that neither the Migdal approximation \cite{Mig} nor 
the usual small-polaron approximation is in quantitative agreement 
with the exact results for intermediate coupling strength. 
So at present the Holstein model cannot be described by any 
single conventional analytical method for the entire range of the 
coupling strength either in the adiabatic or in antiadiabatic limit. 
In a recent work \cite{DJC} we addressed this problem and considering a 
two-site system we developed a perturbation expansion using 
MLF phonon basis and the results obtained thereby 
are in good agreement with the exact numerical 
results for the entire range of the coupling strength for an 
intermediate value of hopping ($t \sim \omega_0$).   
Subsequently, we presented  
a detailed comparison for the convergence of perturbation corrections 
to the ground state energy and wave function for the same system \cite{JA}
using different phonon 
bases obtained through the LF, modified LF (MLF) 
and modified LF with squeezing transformations 
(MLFS). The results showed that for weak and intermediate 
coupling the pertubation corrections within the MLF and MLFS methods 
are much smaller and the convergence of the perturbation 
expansion is much better compared to the LF method. For strong 
coupling all the methods become equivalent and a good convergence 
in the perturbation expansion for the ground state is achieved.  
In this paper we have studied the convergence of the perturbation  
expansion for the energy and the correlation function for the first 
excited state of a two-site single electron system using different 
phonon bases obtained through the LF, MLF, MLFS, LF with squeezing (LFS) 
transformations, and also within weak-coupling expansion. 

In Sec. II we define the model Hamiltonian, describe different 
variational phonon bases states that we have considered and  
present the 
expressions for the energy, wave function and static correlation 
functions calculated within the perturbation method for the first 
excited state. In Sec. III we present the results obtained by 
different methods and discuss about the convergence of the 
perturbation series, hence the applicability of the methods 
in different regions of the $e$-ph coupling strength for different 
hopping parameters and conclusions. 
\vskip 0.2cm
\begin{center}
{\bf II. FORMALISM }
\end{center}
\vskip 0.2cm

The two-site single-polaron Holstein Hamiltonian is
\begin{eqnarray}
H = \sum_{i,\sigma} \epsilon n_{i \sigma} - \sum_{\sigma}
t (c_{1 \sigma}^{\dag} c_{2 \sigma} + c_{2 \sigma}^{\dag} c_{1 \sigma})
+ g \omega_0  \sum_{i,\sigma}  n_{i \sigma} (b_i + b_i^{\dag}) 
+  \omega_0 \sum_{i}  b_i^{\dag} b_i ~ , 
\end{eqnarray} 
where $i$ =1 or 2, denotes the site. $c_{i\sigma}$ ($c_{i\sigma}^{\dag}$)  
is the annihilation (creation) operator for the electron with spin 
$\sigma$ at site $i$, and $n_{i \sigma}$ (=$c_{i\sigma}^{\dag} c_{i\sigma}$) 
is the corresponding number operator. $g$ denotes the on-site $e$-ph coupling  
strength, $t$ is the usual hopping integral. $b_i$ and $b_{i}^{\dag}$ are the annihilation and 
creation operators, respectively, for phonons corresponding to 
interatomic vibrations at site $i$, and $\omega_0$ is the phonon 
frequency. For the one-electron case spin index is redundant 
here, hence omitted in the following. 

The Hamiltonian (1) can be divided into two independent parts: one part 
describing  
symmetric in-phase vibrations coupled to the total number of electrons 
and the other asymmetric out of phase vibrations coupled to the  
electronic degrees of freedom. The first part describes just a 
shifted harmonic oscillator while the solution of the second part 
is a nontrivial problem \cite{RT,JA}. The Hamiltonian for the second part is 
given by
 \begin{eqnarray}
H_d = \sum_{i} \epsilon n_{i} - t (c_{1}^{\dag} c_{2} + c_{2}^{\dag} c_{1}) 
+ \omega_0  g_{+} (n_1-n_2) (d + d^{\dag}) 
+  \omega_0  d^{\dag} d ~,
\end{eqnarray}
where $g_{+}=g/\sqrt 2$ and $d=(b_1 -b_2)/\sqrt{2}$.
 
 For a perturbation method it is desirable to use a basis 
where the major part of the Hamiltonian becomes diagonal. 
The phonon bases chosen in the MLF or MLFS approach consist of 
variational displacement oscillators and can produce retardation 
effect even in the  
zeroth order of perturbation \cite{DS,DC} unlike the LF approach, 
hence these bases are better choices for 
perturbative calculation when the coupling strength is not very strong. 
We now use the MLF transformation where the lattice deformations 
produced by the electron are treated as variational parameters 
\cite {LS,DS,DC}. For the present system, 
\begin{equation}
\tilde{H_d} = e^R H_d e^{-R}~,
\end{equation}
where $R =\lambda (n_1-n_2) ( d^{\dag}-d)$, and $\lambda$ is a
variational parameter related to the displacement of the $d$ 
oscillator. 
The transformed Hamiltonian is then obtained as 
\begin{eqnarray}
\tilde{H_d}&=&\omega_0  d^{\dag} d + \sum_{i} \epsilon_p n_{i} - 
t [c_{1}^{\dag} c_{2}~ \exp(2 \lambda (d^{\dag}-d))    \nonumber\\ 
&+& c_{2}^{\dag} c_{1}~\exp(-2 \lambda (d^{\dag}-d))]  
+ \omega_0  (g_{+} -\lambda) (n_1-n_2) (d + d^{\dag}) ~,
\end{eqnarray}
where $\epsilon_p = \epsilon - \omega_0 ( 2 g_{+}-\lambda)\lambda$. 

Now we will make a squeezing transformation \cite{HZ} to Hamiltonian (4)  
\begin{eqnarray} 
\tilde{H_{sd}} &=& e^S \tilde{H_{d}} e^{-S}   
\end{eqnarray} 
where $S=\alpha(d_id_i-d^{\dag}_{i} d^{\dag}_{i})$. This new phonon 
basis is squeezed with respect to the previous basis. 
The transformed Hamiltonian (5) takes the form 
\begin{eqnarray} 
\tilde{H_{sd}} &=& \omega_0  d^{\dag} d [{\cosh}^2(2\alpha)
 +{\sinh}^2(2\alpha)] + \omega_0~\cosh(2\alpha)~\sinh(2\alpha)  
(dd+ d^{\dag}d^{\dag})  \nonumber\\  
&+& \sum_{i} \epsilon_p n_{i} 
- t [c_{1}^{\dag} c_{2}~\exp(2 \lambda_{e} (d^{\dag}-d))    
+ c_{2}^{\dag} c_{1}~\exp(-2 \lambda_{e} (d^{\dag}-d))] \nonumber\\  
&+& \omega_0  (g_{+} -\lambda) (n_1-n_2) (d + d^{\dag})~\exp(2 \alpha)
+\omega_0~{\sinh}^2(2\alpha)  
\end{eqnarray} 
where $\lambda_{e}=  \lambda ~e^{-2\alpha} $. 
For the single polaron problem we choose the basis set 
\begin{equation}
|\pm,N \rangle = \frac{1}{\sqrt 2} (c_{1}^{\dag} \pm  c_{2}^{\dag}) 
|0\rangle_e  |N\rangle 
\end{equation}
where $|+\rangle$ and $|-\rangle$ are the bonding and antibonding 
electronic states and $|N\rangle$ denotes the $N$th excited oscillator  
state in the MLFS, MLF, LFS or LF bases depending on the method considered. 
It may be noted that the MLFS basis turns into the MLF basis when 
$\alpha =0$, and into the LFS basis if $\lambda=g_+$. The 
MLF basis turns into the LF basis when $\lambda= g_+$, and it turns 
into the weak coupling expansion for $\lambda= 0$.
We consider the diagonal part of Hamiltonian (6) as the 
unperturbed Hamiltonian ($H_0$) and the remaining part of the 
Hamiltonian, $H_{1}= \tilde{H_{sd}}-H_0$, as a perturbation 
\cite{DJC,JA}. 
The unperturbed energy of the state $| \pm,N\rangle$ is given by 
\begin{eqnarray}
 E_{\pm,N}^{(0)}= \langle N,\pm|H_0|\pm, N \rangle=
 \omega_0 [{\sinh}^2(2\alpha) + N({\sinh}^2(2\alpha) 
 + {\cosh}^2(2\alpha))]\nonumber\\
 + \epsilon_p \mp t_{e} \left[ \sum_{i=0}^{N}
 (\frac{(2\lambda_{e})^{2i}}{i!} (-1)^i N_{C_i}\right]
\end{eqnarray}
where $t_{e}=t~\exp{(-2\lambda_{e}^2)}$.
The general off-diagonal matrix elements of 
$H_1$ are given in Refs. \cite{DJC,JA}.
As noted in our previous work \cite {DJC} the unperturbed first excited state 
wave function should be built up as a linear combination of 
$|+,1\rangle$ and $|-,0\rangle$. 
The unperturbed energies of the states $|+,1\rangle$ and 
$|-,0\rangle$ are ($\epsilon_p +\omega_0 -t_e(1-4\lambda^2)$ 
$+3{\sinh}^2(2\alpha))$ and 
($\epsilon_p +t_e+\omega_0 {\sinh}^2(2\alpha)$), 
respectively within the MLFS approach, and the off-diagonal matrix 
element between these two states is 
($2\lambda_e t_e+\omega_0(g_{+}-\lambda) e^{2\alpha}$). 
The unperturbed energies of these two states cross at an 
intermediate value of $g_{+}$ for $2t > \omega_0$. So, following  
degenarate perturbation theory linear combinations of  
the states $|+,1\rangle$ and $|-,0\rangle$ are formed to obtain 
two new elements of basis states so that $\tilde{H_{sd}}$ becomes 
diagonal in the sub-space spanned by these two states.
 The unperturbed first excited 
 state is given by 
\begin{eqnarray}  
|\psi_1^{(0)}\rangle= \frac{1}{\sqrt{(a^2+b^2)}}
\left[ a |-,0\rangle  + b|+,1\rangle \right]
\end{eqnarray}  
The ratio ($c$) of the coefficients $b$ to $a$ and the unperturbed energy  
($\alpha$) of the first excited state may be found out from the relation 
\begin{eqnarray}  
c = \frac{\alpha-H_{11}}{H_{12}}
=\frac{H_{12}}{\alpha-H_{22}}
\end{eqnarray}  
where $H_{11}$, $H_{22}$, $H_{12}$ are the matrix elements of $\tilde{H_{d}}$
 in the subspace of $|-,0\rangle$ and 
$|+,1\rangle$.
Eq.(10) gives two roots of $\alpha$, the lower value of $\alpha$ 
(say, $\alpha_1$) corresponds to the first excited state. 

Our chosen phonon basis within the MLFS method have two variational 
parameters $\lambda$ and $\alpha$, proper choices of 
them will make the perturbative expansion convergent.   
In our previous works \cite {DJC,JA} we find that the minimization of the 
unperturbed ground state energy of the system yields phonon 
bases (within the MLF and MLFS) for which the perturbative expansion for the 
ground state shows satisfactory convergence.  
The corresponding values of $\lambda$ and $\alpha$ are given by 
$\lambda =\frac {\omega_0g_{+}}{\omega_0+2t_{e} e^{-4\alpha}}$ and  
$\alpha = \frac{1}{4}{\sinh}^{-1}[2\lambda(g_+-\lambda)] $. 
For the study of the first excited state we will use the same basis 
as we have used for the ground state. 
The first-order correction to the first excited state wave function 
is obtained as,
\begin{eqnarray}  
|\psi_1^{(1)}\rangle &=&\frac{1}{\sqrt{1+c^2}} \left[ \sum_{N=2,4,..}
\frac{W_e}{(\alpha_1-E_{-,N}^{(0)})}~~ |-,N\rangle \right. \nonumber\\
&+&\left. \sum_{N=3,5..}\frac{W_o}{( \alpha_1-E_{+,N}^{(0)})}
~~|+,N\rangle  \right]
\end{eqnarray}  
where $W_e= W +\sqrt{N}~ \omega_0 c(g_{+}-\lambda)~e^{2\alpha}~\delta_{N,2}$
 + $\sqrt{N(N-1)}~\frac{\omega_0}{2}~\sinh(4\alpha)~\delta_{N,2}$, \\ 
$W_o=W+\sqrt{N} \omega_0 (g_{+}-\lambda) e^{2\alpha}\delta_{N,1}$
+ $\sqrt{N(N-1)}~\frac{\omega_0}{2}~c~{\sinh}(4\alpha)~\delta_{N,3}$, \\
and $ W=t_e\frac{(2\lambda_e)^N}{\sqrt{N!}}(1+2\lambda_e c-\frac{cN}
{2\lambda_e})$. 

Second-order correction to the first excited state energy is given by,
\begin{eqnarray}
E_1^{(2)} = \frac{1}{1+c^2}\left[ \sum_{N=2,4,..}
\frac{|W_e|^2}{(\alpha_1-E_{-,N}^{(0)})} 
+ \sum_{N=3,5..}\frac{|W_o|^2}{( \alpha_1-E_{+,N}^{(0)})}  \right]
\end{eqnarray}
Using our previous prescription \cite {JA} successive 
higher order corrections to the 
wave function ($|\psi_1^{(N)}\rangle$) as well as to the 
energy ($E_1^{(N)}$) are calculated.  
   
   The static correlation functions $\langle n_1 u_{1}\rangle_{0}$ and 
$\langle n_1 u_{2}\rangle_{0}$,
where $u_1$ and $u_2$ are the lattice deformations at sites 1 and 2, 
respectively, produced by an electron at site 1, are the standard 
measure of polaronic character, and indicate the strength of polaron 
induced lattice deformations and their spread. The operators 
involving the correlation functions may be written as 
\begin{equation}  
n_{1} u_{1,2} = \frac{n_1}{2}[(a+a^{\dag})\pm(d+d^{\dag})
e^{2\alpha} -~2~(n g_+\pm \lambda(n_1-n_2))]\nonumber \\
\end{equation}
The final form of the correlation functions are obtained as
\begin{eqnarray}
\langle n_{1} u_{1} \rangle_{0}&=&\frac{1}{2}  
\left[-(g_{+} +\lambda) + \frac{A_1~e^{2\alpha}}{N_{G1}}\right]~, \\
\langle n_{1} u_{2} \rangle_{0}&=&\frac{1}{2}
\left[-(g_{+}-\lambda)- \frac{A_1~e^{2\alpha}}{N_{G1}}\right] ~,\nonumber
\end{eqnarray}  
where $N_{G1} = \langle \psi_{1}|\psi_{1}\rangle $, 
$A_1 = \langle \psi_{1} |n_1(d+d^{\dag})|\psi_{1}\rangle$ and 
$|\psi_{1}\rangle$ is the perturbed wave function of the first excited 
state.

\vskip .2cm
\begin{center}
{\bf IV. RESULTS AND DISCUSSIONS}
\end{center}
\vskip 0.2cm

The perturbation 
corrections to the energy and wave function of the first excited state 
are estimated up to the sixth and fifth-orders, respectively, 
within the LF, LFS, MLF, MLFS methods, and weak-
coupling perturbative method considering 25 phonon states 
(which is sufficient for $g_+\leq ~2.2$) in the transformed 
phonon basis. 

 In Fig. 1 the perturbation corrections to the energy 
of the first excited state are plotted 
as a function of $g_+$ for $t/\omega_0=1.1$. The perturbation 
corrections in energy 
are smaller within the MLF and MLFS than the LFS and LF methods. 
For the LF, LFS, and MLFS methods the convergence of the corrections 
is weaker in a range of $g_+$ where third and fourth or 
fifth and sixth-order corrections are comparable. Within MLF approach 
energy perturbation corrections show satisfactory 
convergence in the entire region of $g_+$ and the energy of 
the first excited state, when computed considering  
up to the fifth or sixth-order corrections, becomes identical 
with the exact result of Ref. \cite{RT}. 

 The shape of the correlation function $\langle n_{1} u_{2} \rangle_{0}$
 is much more sensitive to the corrections to the wave function  than 
 $\langle n_{1} u_{1} \rangle_{0}$ which indicates that convergence of 
 $\langle n_{1} u_{2} \rangle_{0}$ is a clear signature for convergence 
 of the wave function. In Fig. 2 we plot 
the correlation function 
$\langle n_{1} u_{2} \rangle_{0}$ obtained by considering up to the 
different orders of perturbation corrections to the wave function 
against $g_+$. 
The LF method shows a bad convergence for low values of $g_+$, 
while a good convergence beyond $g_+=1.0$. 
The convergence within the LFS method is not satisfactory over
a wide region of $g_+$. 
The MLF method shows very good convergence for low 
values of $g_+$ ($\le 1.0$) as well as for high values of $g_+$ 
($\ge 1.3$). 
Convergence within the MLFS is excellent for low values of 
$g_+$ , but then for a wide region of intermediate values of 
$g_+$ the convergence is not satisfactory. 
When compared with exact results of 
$\langle n_{1} u_{2} \rangle_{0}$ (taken from the Ref.\cite{RT}), 
it is found that the MLF results up to the fifth order perturbation 
are identical with the exact results except in the region 
$0.9 < g_+<1.3$ where a slight departure in values from the exact 
results is seen. 

In Fig. 3 we plot the perturbation corrections of different orders 
to the first excited state energy and the 
the correlation function 
$\langle n_{1} u_{2} \rangle_{0}$ evaluated by considering up to  
different orders of perturbation corrections to the wave function 
against $g_+$ within the weak-coupling perturbative expansion 
for $t/\omega_0=1.1$. 
The odd order corrections to the energy within 
the weak-coupling expansion are zero. The energy 
corrections of any (even) order increases monotonically 
with $g_+$, since the entire $e$-ph interaction is treated as  
a perturbation within the weak-coupling scheme. The correlation 
function within the weak-coupling perturbation procedure 
shows good convergence for low values of $g_+$, as expected. 
However even in this region, convergence within the 
MLF and MLFS methods 
is found to be slightly better than that obtained within 
the weak-coupling scheme. 

In Fig. 4 we show the energy corrections and correlation function, 
calculated by considering different order of corrections, within the 
MLF method for $t/ \omega_0=0.6$. The figure shows that the 
perturbation corrections of successive orders are really very small  
and shows perfect convergence for the entire region of $g_+$. 
The correlation function calculated considering up to second order 
correction to the wave function would reproduce almost the exact 
result in this limit. For lower values of $t$ obviously the 
convergence within the MLF scheme would be much better. 
We have also studied the perturbation corrections to the energy 
and the correlation function $\langle n_{1} u_{2} \rangle_{0}$ 
for the first excited state in the adiabatic region of hopping 
($t/ \omega_0=2.1$) within MLF approach and found that 
energy convergence is quite good for 
the entire range of $g_+$ whereas 
convergence of corrections to the wave function is fairly well 
except for intermediate range of coupling.

In conclusion, our study on the first excited state of a two-site 
single electron Holstein model shows that the perturbation method 
based on the MLF variational phonon basis is better than the 
other methods (LF, LFS, MLFS) if the entire range of $g_+$ 
is considered. 
The MLF method could yield exact results 
for the entire range of $g_+$ in the antiadiabatic cases 
($t/\omega_0 \le 0.6$) and for a major region of $g_+$ (covering 
both low and high values of $g_+$) 
for intermediate hopping ($t/\omega_0=1.1$). 
For small values of $g_+$ both the MLFS and MLF methods show excellent 
convergence, convergence is found to be even better than that within 
the weak-coupling perturbation  
method. 
Our previous study on the ground state and the present study on the first 
excited state thus establish that the perturbation method based on the 
MLF phonon basis could 
satisfactorily describe the ground state as well as the first excited 
state of the two-site system for major region of the $e$-ph coupling 
strength.

\newpage
Figure captions :

\noindent
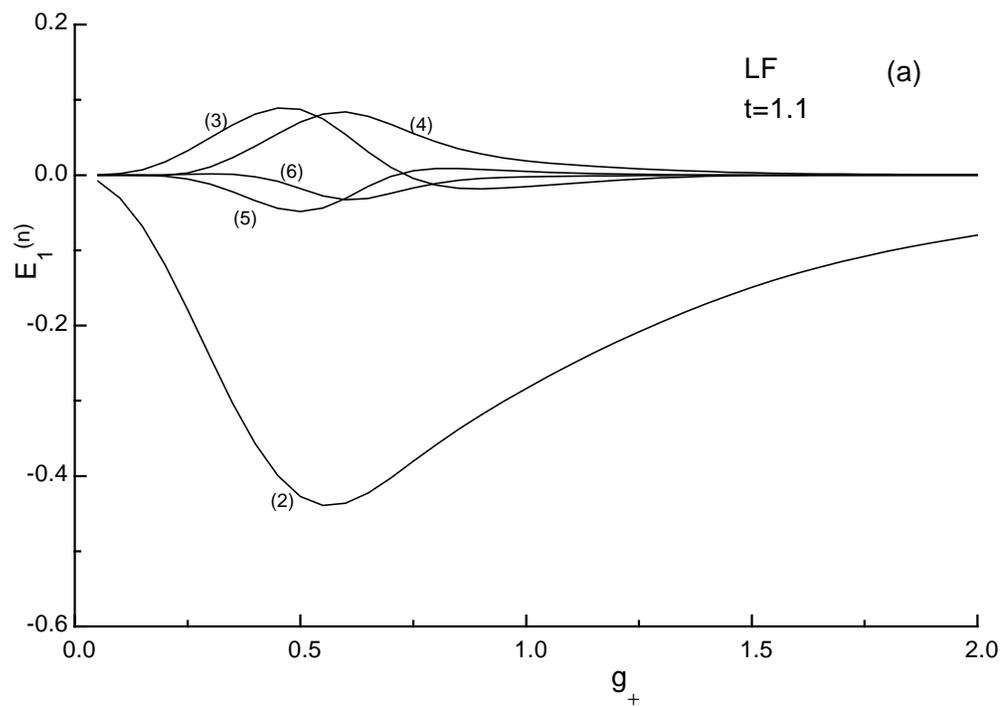
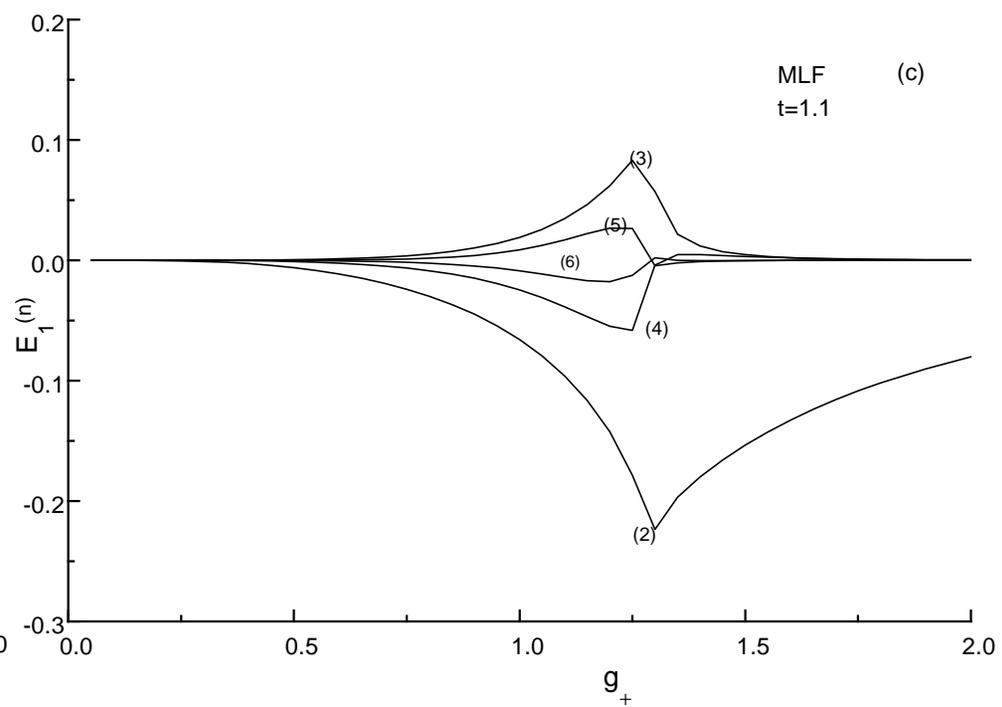
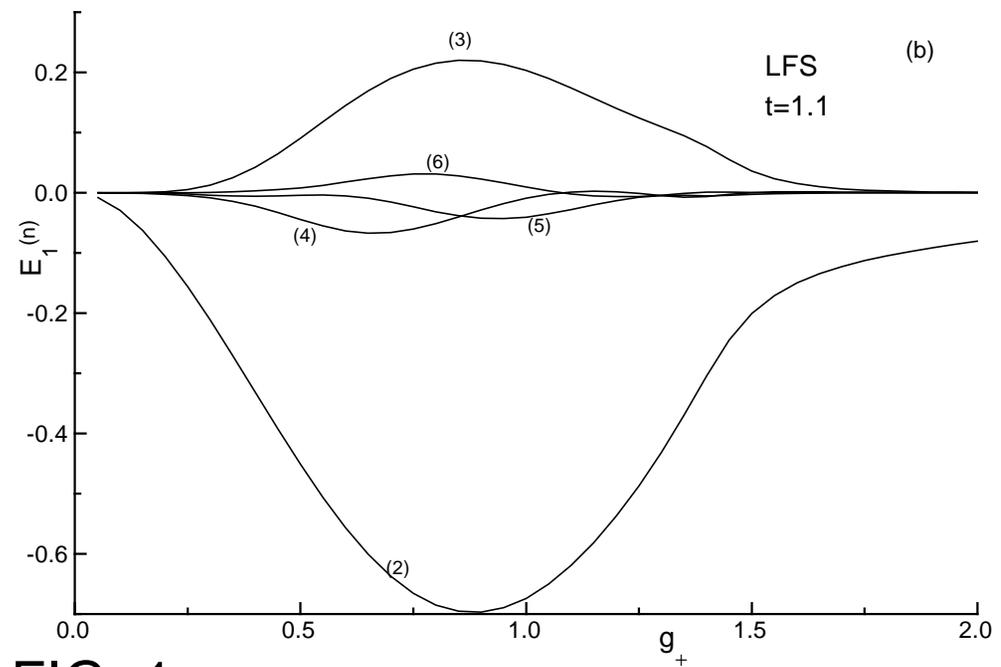
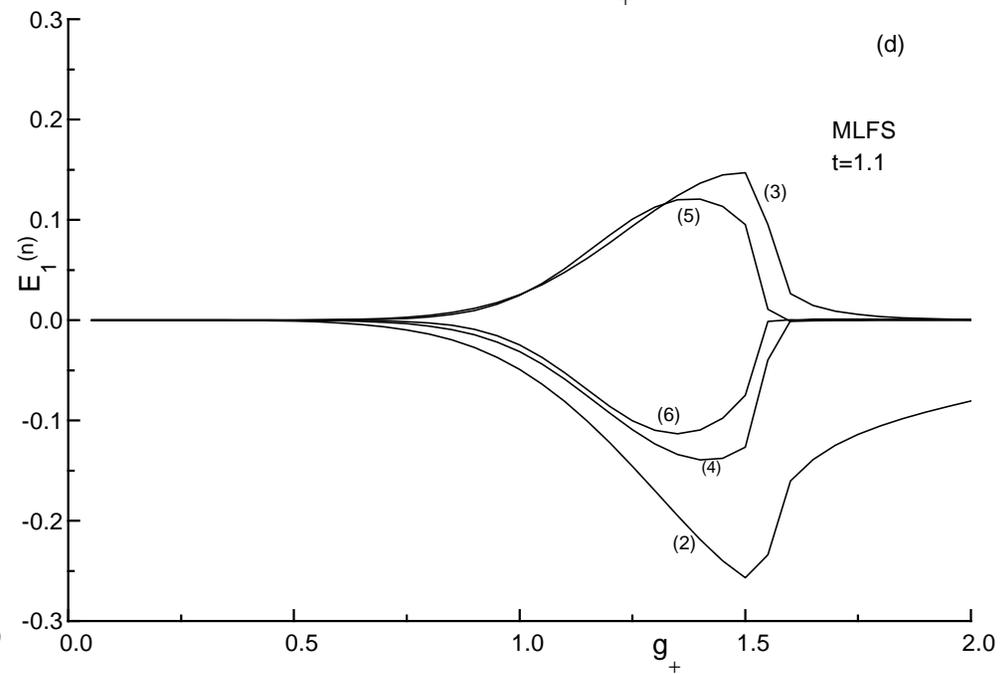
FIG. 1. Variation of the perturbation corrections 
$E_{1}^{(n)}$ to the first excited state energy as a 
function of the coupling strength ($g_{+}$) for $t/\omega_0 =1.1$ 
in (a) LF, (b) LFS, 
(c) MLF, and (d) MLFS methods. 
$E_{1}^{(n)}$ is the $n$th order perturbation correction to the 
first excited state energy. 

\vskip 0.5 cm
\noindent

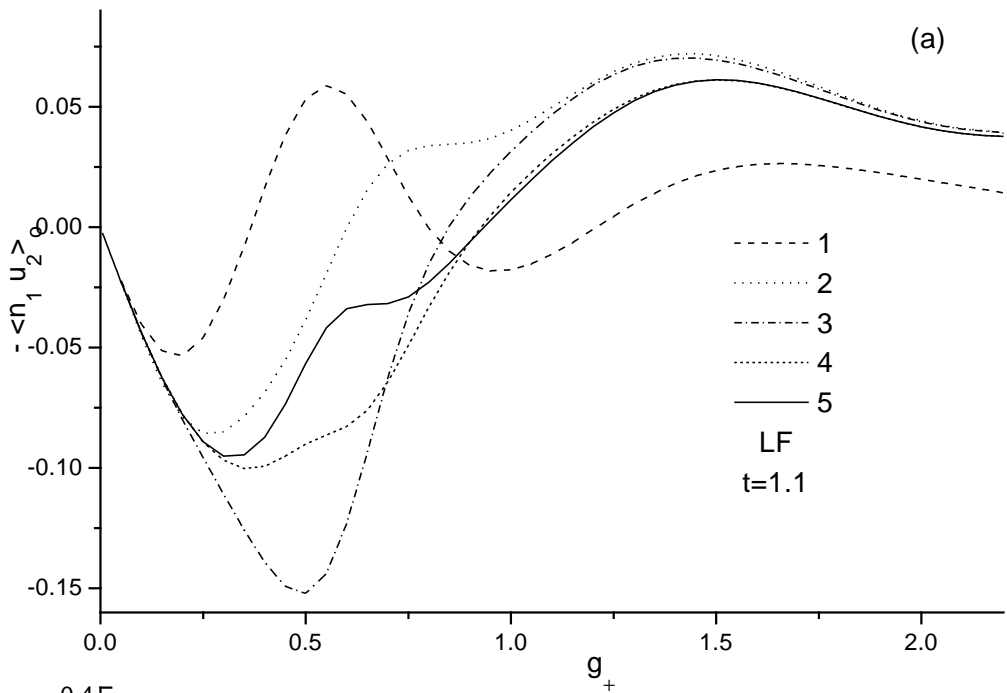
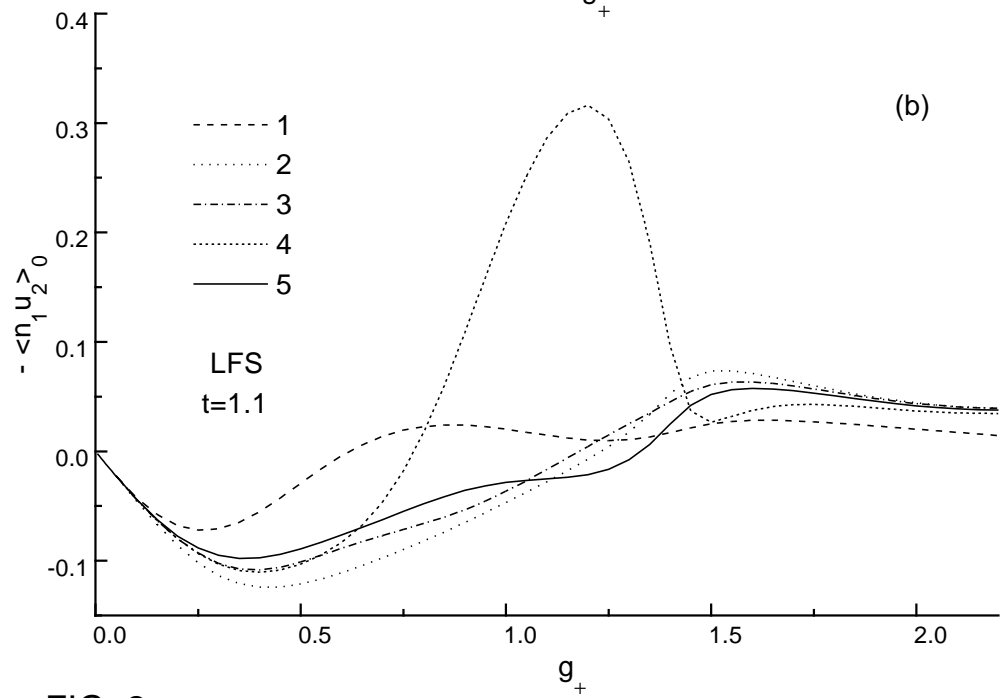
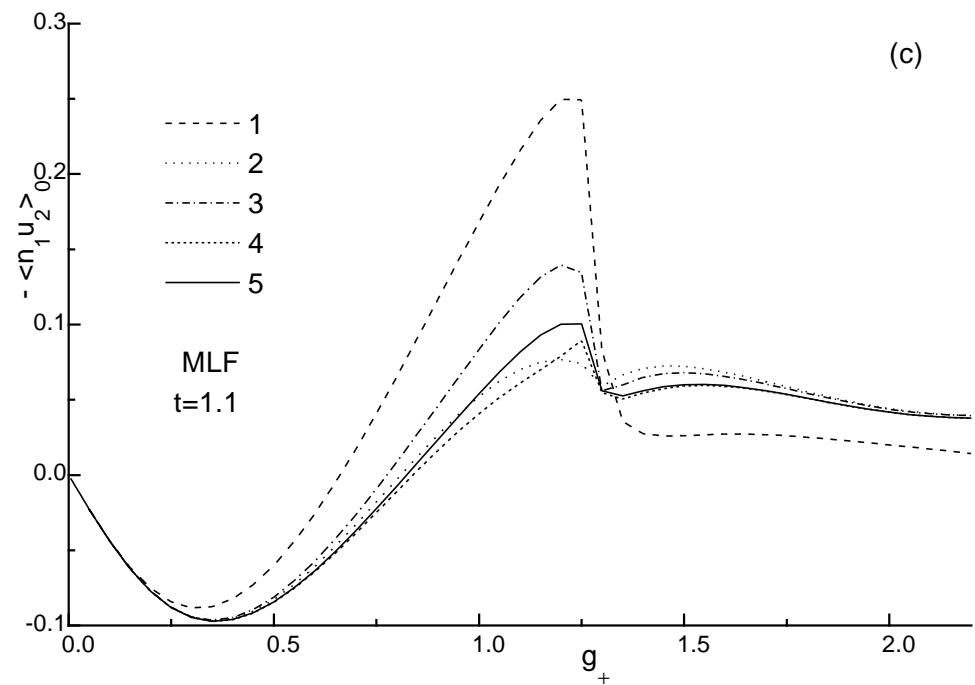
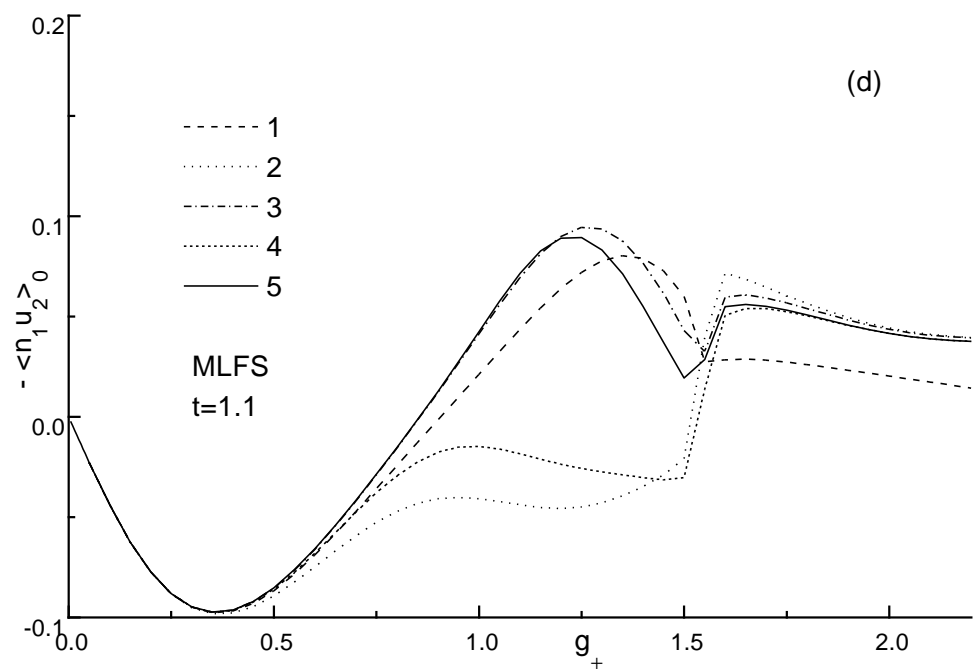
FIG. 2. Plot of the correlation function $\langle n_1u_2 \rangle_{0}$ 
calculated up to different order of perturbations in the wave function 
vs $g_{+}$ for $t/\omega_0 =1.1$ 
within different methods (a) LF, (b) LFS, 
(c) MLF, and (d) MLFS. 
The labels (2), (3), .... denote the curves obtained by considering  
up to the second-order, third-order, .... corrections 
to the wave function, 
respectively.
\vskip 0.5 cm     
\noindent

\vskip 0.5 cm
\noindent

FIG. 3 Convergence within weak-coupling expansion for $t/\omega_0 =1.1$ : 
(a) the perturbation corrections to the first excited state energy and 
(b) correlation function $\langle n_1u_2 \rangle_{0}$ 
calculated up to different order of perturbations in the wave function, 
as a function of $g_{+}$. 
\vskip 0.5 cm     
\noindent

FIG. 4 Convergence within MLF approach for $t/\omega_0 =0.6$ : 
(a) the perturbation corrections to the first excited state energy and 
(b) correlation function $\langle n_1u_2 \rangle_{0}$ 
calculated up to different order of perturbations in the wave function, 
as a function of $g_{+}$. 
\vskip 0.5 cm     
\noindent
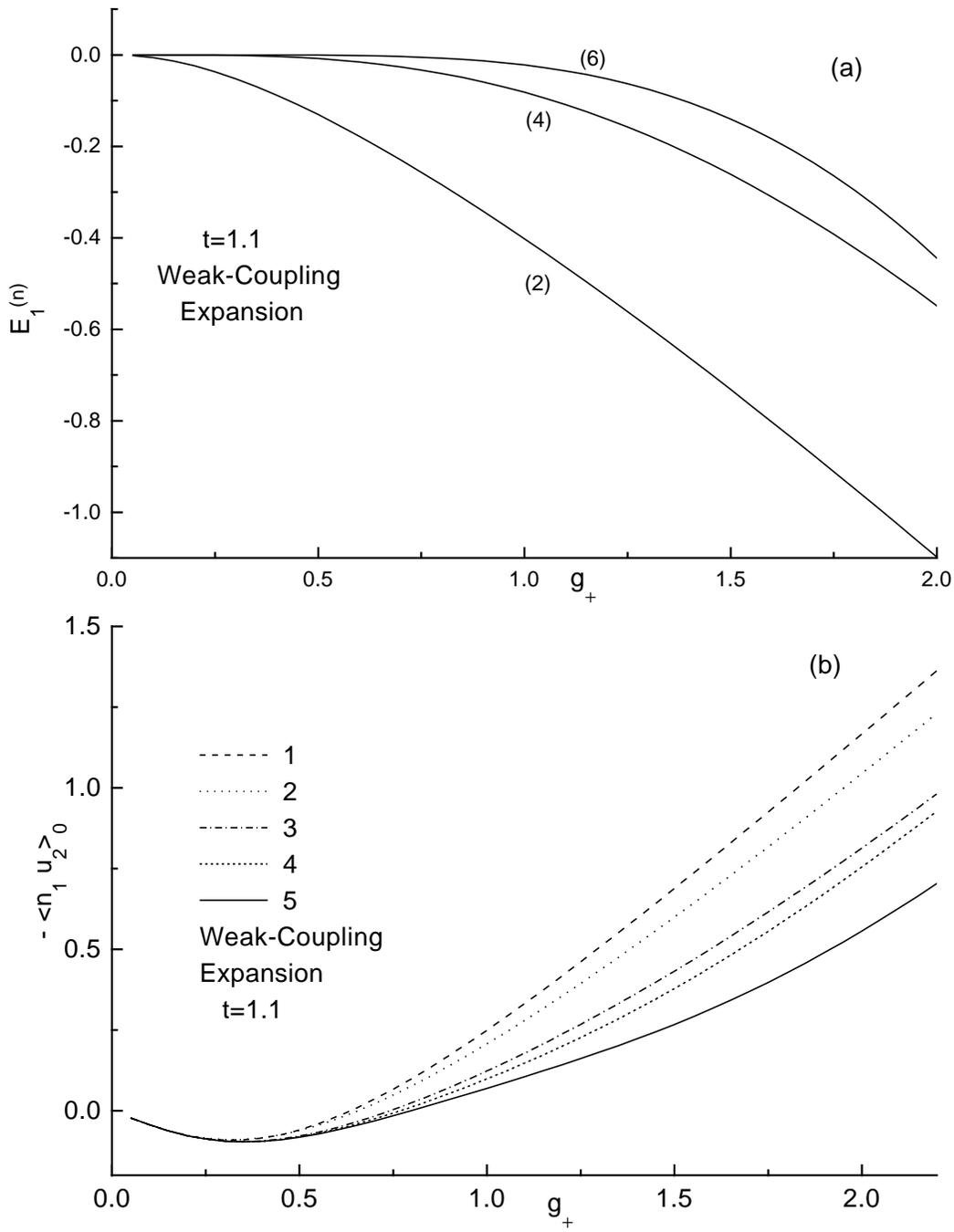
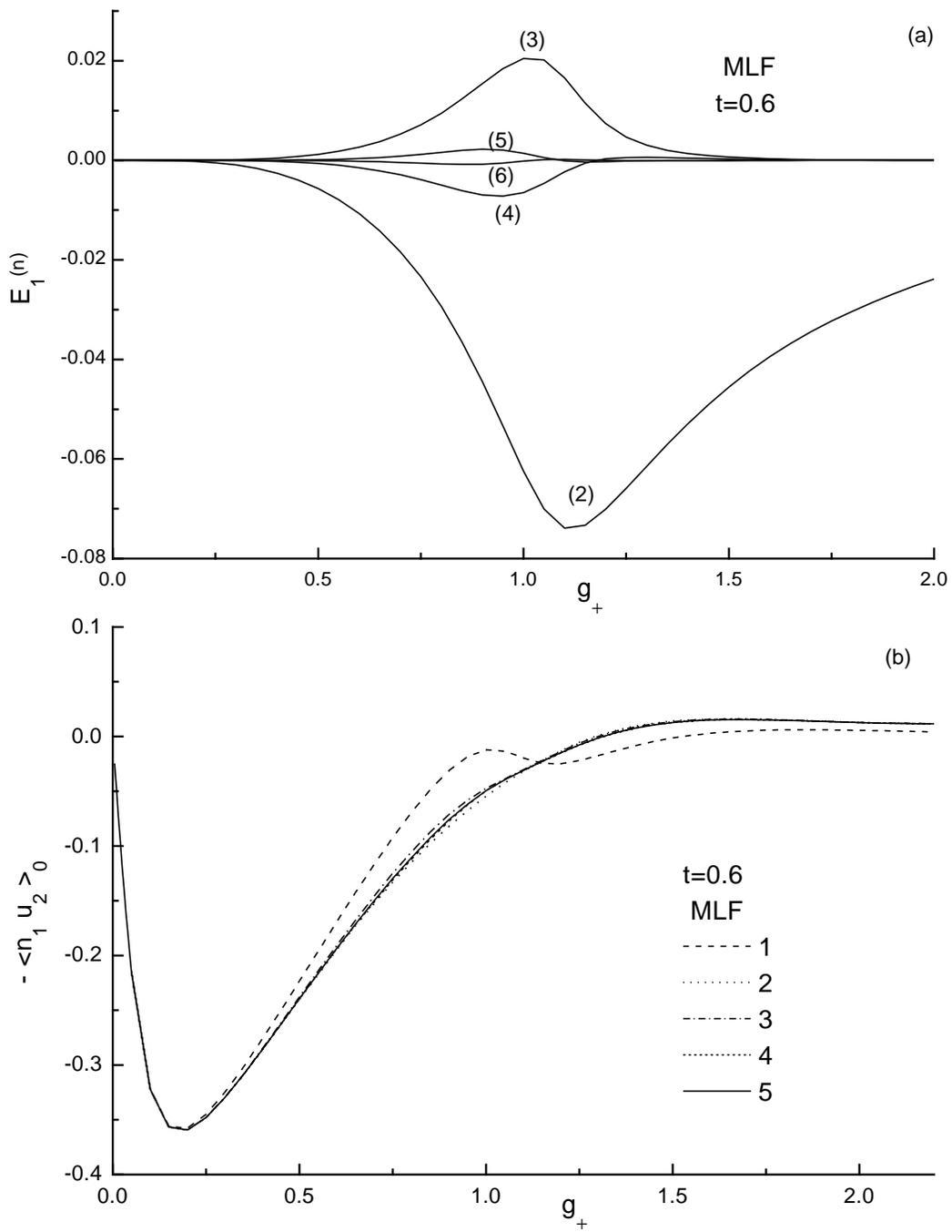
\end{document}